# Subgraph Centrality in Complex Networks

Ernesto Estrada[1*] and Juan A. Rodríguez-Velázquez[2]

[1]Complex Systems Research Group, X-Rays Unit, RIAIDT, Edificio CACTUS, University of Santiago de Compostela, 15706 Santiago de Compostela, Spain and [2]Department of Mathematics, University Carlos III de Madrid, 28911 Leganés (Madrid), Spain.

---

[*] To whom reprint request should be addresses. E-mail: estrada66@yahoo.com

We introduce a new centrality measure that characterizes the participation of each node in all subgraphs in a network. Smaller subgraphs are given more weight than larger ones, which makes this measure appropriate for characterizing network motifs. We show that the subgraph centrality (*SC*) can be obtained mathematically from the spectra of the adjacency matrix of the network. This measure is better able to discriminate the nodes of a network than alternate measures such as degree, closeness, betweenness and eigenvector centralities. We study eight real-world networks for which *SC* displays useful and desirable properties, such as clear ranking of nodes and scale-free characteristics. Compared with the number of links per node, the ranking introduced by *SC* (for the nodes in the protein interaction network of *S. cereviciae*) is more highly correlated with the lethality of individual proteins removed from the proteome.

# I. INTRODUCTION

Complex networks, consisting of sets of nodes or vertices joined together in pairs by links or edges, appear frequently in various technological, social and biological scenarios [1–5]. These networks include the Internet [6], the World Wide Web [7], social networks [8–10], scientific collaboration networks [11], lexicon or semantic networks [12,13], neural networks [14], food webs [15], metabolic networks [16], and protein–protein interaction networks [17]. They have been shown to share global statistical features, such as the "small world" and the "scale-free" effects, as well as the "clustering" property. The first feature is simply the fact that the average distance between nodes in the network is short and usually scales logarithmically with the total number of nodes [18]. The second is a characteristic of several "real-world" networks in which there are many nodes with low degree and only a small number with high degree (the so-called "hubs" [19]). The node degree is simply the number of ties a node has with other nodes. In scale-free networks, the node degree follows a power-law distribution. Finally, clustering is a property of two linked nodes that are each linked to a third node [7]. In consequence, these three nodes form a triangle and the clustering is frequently measured by counting the number of triangles in the network [20].

It has been observed that not only triangles but also other subgraphs are significant in real networks. We say that a graph $G'=(V',E')$ is a subgraph of $G=(V,E)$ if $V' \subseteq V$ and $E' \subseteq E$. The term "network motifs" designates those patterns that occur in the network far more often than in random networks with the same degree sequence [21]. Network motifs found in technological and biological networks are small subgraphs that capture specific patterns of interconnection characterizing the networks at the local level [21,22].

# II. CENTRALITY MEASURES

Another kind of local characterization of networks is made numerically by using one of several measures known as "centrality" [23]. One of the most used centrality measures is the "degree centrality", *DC* [7], which is a fundamental quantity describing the topology of scale-free networks [18]. *DC* can be interpreted as a measure of immediate influence, as opposed to long-term effect in the network [23]. For instance, if a certain proportion of nodes in the network are infected, those nodes having a direct connection with them will also be infected. However, although a node in a network may be linked to only one node, the risk of infection to the first node remains high if the latter is connected to many others.

There are several other centrality measures that have been introduced and studied for real world networks, in particular for social networks. They account for the different node characteristics that permit them to be ranked in order of importance in the network. Betweenness centrality (*BC*) characterizes how influential a node is in communicating between node pairs [24]. In other words, *BC* measures the number of times that a shortest path between nodes i and j travels through a node k whose centrality is being measured. The farness of a vertex is the sum of the lengths of the geodesics to every other vertex. The reciprocal of farness is closeness centrality (*CC*). The normalized closeness centrality of a vertex is the reciprocal of farness divided by the minimum possible farness expressed as a percentage [7,23]. This measure is only applicable to connected networks, since the distance between unconnected nodes is undefined. Neither *BC* nor *CC* can be related to the network subgraphs in a way that permits them to be considered as measures of node subgraph centrality.

A centrality measure that is not restricted to shortest paths is the eigenvector centrality (*EC*) [25], which is defined as the principal or dominant eigenvector of the adjacency matrix A representing the connected subgraph or component of the network. It simulates a mechanism in which each node affects all of its neighbors simultaneously [26]. *EC* cannot be considered as a measure of centrality whereby nodes are ranked according to their participation in different network subgraphs. For instance, in a graph with all nodes having the same degree (a regular graph), all the components of the main eigenvalue are identical [27], even if they participate in different subgraphs. *EC* is better interpreted as a sort of extended degree centrality which is proportional to the sum of the centralities of the node' neighbors. Consequently, a node has high value of *EC* either if it is connected to many other nodes or if it is connected to others that themselves have high *EC* [28].

In Fig. 1, we illustrate two regular graphs, with eight and nine nodes, and degrees equal to 3 and 6, respectively. In graph (a), nodes $\{1,2,8\}$ are the only ones forming part of a triangle. Vertices $\{4,6\}$ form part of three squares, vertices $\{3,5,7\}$ form part of only two and the rest do not form part of any. The analysis can be obviously extended to larger subgraphs. However, it is evident that there are three groups of distinguishable vertices in the graph: $\{1,2,8\}$, $\{4,6\}$ and $\{3,5,7\}$. These are distinguishable according to their participation in the different subgraphs, although they cannot be distinguished by *EC*. In graph (b), vertices $\{1,3,5,6,8\}$ take part in 44 of the 100 squares present in the graph, while vertices $\{2,4,7,9\}$ take part in 45 (all vertices take part in the same number of smaller subgraphs; e.g., edges, triangles, connected triples). However, these groups of vertices cannot be distinguished by any of the centrality measures (*DC*, CC, *BC* and *EC*).

Insert Fig. 1 about here.

In this work, we propose a method for characterizing nodes in a network according to the number of closed walks starting and ending at the node. Closed walks are appropriately weighted such that their influence on the centrality decreases as the order of the walk increases. Each closed walk is associated with a connected subgraph, which means that this measure counts the times that a node takes part in the different connected subgraphs of the network, with smaller subgraphs having higher importance. Consequently, we will call this measure the "subgraph centrality" (*SC*) for nodes in a network.

**III. SUBGRAPH CENTRALITY MEASURE**

Let $G$ be a simple graph of order $N$. The graph spectrum is the set of eigenvalues of the adjacency matrix of the graph. Graph spectral density is the density of the eigenvalues of its adjacency matrix, which can be directly related to the topological features of the graph through the spectral moments [29,30]. For instance, the number of closed walks of length $k$ starting and ending on vertex $i$ in the network is given by the local spectral moments $\mu_k(i)$, which are simply defined as the $i$th diagonal entry of the $k$th power of the adjacency matrix, **A**:

$$\mu_k(i) = \left(\mathbf{A}^k\right)_{ii} \tag{1}$$

These closed walks are directly related to the subgraphs of the network. For instance, a closed walk of order three represents a triangle, closed walks of order four represent subgraphs of four nodes, and so forth. It is worth noting to comment that even closed walks, i.e., those going back and forth through an even number of edges, can be trivial. A trivial closed walk is that describing a subgraph that does not contains any cycle, i.e., acyclic subgraphs. In Table 1 we illustrate the closed walks of length four (two trivial and one non-trivial) and the subgraphs described by them.

Insert Table 1 about here.

We define the subgraph centrality of the vertex $i$ as the "sum" of closed walks of different lengths in the network starting and ending at vertex $i$. As this sum includes both trivial and non-trivial closed walks we are considering all subgraphs, i.e., acyclic and cyclic, respectively. The contribution of these closed walks decreases as the length of the walks increases. That is, shorter closed walks have more influence on the centrality of the vertex than longer closed walks. This rule is based on the observation that motifs in real-world networks are small subgraphs. The extreme case is that of closed walks of length two only, giving a weight of zero to longer walks. This case corresponds to the vertex degree centrality. On the other hand, the use of the sum of closed walks for defining subgraph centrality presupposes a mathematical problem as the series $\sum_{k=0}^{\infty} \mu_k(i) = \infty$ diverges. Consequently, we avoid this problem by scaling the contribution of closed walks to the centrality of the vertex by dividing them by the factorial of the order of the spectral moment. That is, the *subgraph centrality* of vertex $i$ in the network is given by:

$$SC(i) = \sum_{k=0}^{\infty} \frac{\mu_k(i)}{k!}. \tag{2}$$

Let $\lambda$ be the main eigenvalue of **A**. For any nonnegative integer $k$ and any $i \in \{1,...,n\}$, $\mu_k(i) \leq \lambda^k$, series (2), whose terms are nonnegative, converges.

$$\sum_{k=0}^{\infty} \frac{\mu_k(i)}{k!} \leq \sum_{k=0}^{\infty} \frac{\lambda^k}{k!} = e^{\lambda} \tag{3}$$

Thus, the subgraph centrality of any vertex $i$ is bounded above by $SC(i) \leq e^{\lambda}$. The following result shows that the subgraph centrality can be obtained mathematically from the spectra of the adjacency matrix of the network.

**Theorem**: *Let $G = (V, E)$ be a simple graph of order $N$. Let $v_1, v_2, ..., v_N$ be an orthonormal basis of $R^N$ composed by eigenvectors of **A** associated to the eigenvalues $\lambda_1, \lambda_2, ..., \lambda_N$. Let $v_j^i$* denote the $i$th component of $v_j$. For all $i \in V$, the subgraph centrality may be expressed as follows:

$$SC(i) = \sum_{j=1}^{N} \left(v_j^i\right)^2 e^{\lambda_j} \qquad (4)$$

*Proof*: The orthogonal projection of the unit vector $e_i$ (the $i$th vector of the canonical base of $R^n$) on $v_j$ is

$$pr_j(e_i) = \frac{\langle e_i, v_j \rangle}{\|v_j\|^2} v_j = \langle e_i, v_j \rangle v_j = v_j^i \cdot v_j. \qquad (5)$$

Hence, the number of closed walks starting at vertex $i$ can be expressed in terms of the spectral properties of the graph as follows:

$$\mu_k(i) = \left(\mathbf{A}^k\right)_{ii} = \langle \mathbf{A}^k e_i, e_i \rangle = \left\langle \mathbf{A}^k \sum_{j=1}^{N} pr_j(e_i), \sum_{j=1}^{N} pr_j(e_i) \right\rangle = \sum_{j=1}^{N} \lambda_j^k \left(v_j^i\right)^2. \qquad (6)$$

Using expression (2), we obtain

$$SC(i) = \sum_{k=0}^{\infty} \left( \sum_{j=1}^{N} \frac{\lambda_j^k \left(v_j^i\right)^2}{k!} \right). \qquad (7)$$

By reordering the terms of series (7), we obtain the absolutely convergent series:

$$\sum_{j=1}^{N} \left( (v_j(i))^2 \sum_{k=0}^{\infty} \frac{\lambda_j^k}{k!} \right) = \sum_{j=1}^{N} \left( (v_j(i))^2 e^{\lambda_j} \right), \qquad (8)$$

which, obviously, also converges to $SC(i)$. Thus, the result follows.

It has been stated by previous authors that among all graphs with *n* nodes, the maximal centrality should be attained by the hub of a star [31]. A star with *n* nodes, designed as $S_n$, is a tree with one node having degree $n-1$ and the others having degree 1. However, in terms of the number of times a vertex takes part in network subgraphs, the perspective is different. For instance, a vertex in the complete graph $K_n$ takes part in a higher number of subgraphs than the hub of the star $S_n$ (for $n \geq 3$). The complete graph, $K_n$, is the graph of *n* nodes in which each pair of nodes is connected by an edge. $K_n$ can be decomposed into one subgraph isomorphic to $S_n$ and $(n-1)(n-2)/2$ edges, which means that all subgraphs contained in the star $S_n$ are a subclass of the subgraphs contained in the complete graph $K_n$. Take for instance the simple example of $K_5$ and $S_5$. Any node of $K_5$ takes part in 6 connected triples and 2 triangles, which are the only two connected subgraphs of three nodes that exist. However, the central node of $S_5$ takes part in 6 connected triples but each of the other nodes take part in only two and none of these nodes take part in any triangle, showing that nodes in the complete graph take part in a higher number of subgraphs that nodes in the star. In other words, any vertex of $K_n$ takes part in the same number of (acyclic) subgraphs in which the hub of the star participates plus in many other acyclic and cyclic subgraphs. In general, of all connected graphs with *n* nodes, the maximal subgraph centrality is attained by the vertices of the complete graph.

**Proposition:** *Let G be a simple and connected graph of order $n > 1$. Then for every vertex i,*
$$SC(i) \leq \frac{1}{n}\left[e^{n-1} + \frac{n-1}{e}\right].$$
*The equality holds if and only if G is the complete graph $K_n$.*

*Proof*: Since *G* is nontrivial, let *x* be an edge of *G*. Let *G–x* be the graph obtained by removing *x* from *G*. Then the number of closed walks of length *k* in *G–x* is equal to the number of closed walks of length $k$ in $G$ minus the number of closed walks of length k in $G$ containing $x$. Consequently, for all $i$, $SC(i)$ in $G$–$x$ is lower than or equal to $SC(i)$ in $G$. In closing, the maximum $SC(i)$ is attained if and only if $G$ is the complete graph $K_n$.

We now compute $SC(i)$ in $K_n$. The eigenvalues of $K_n$ are $n-1$ and $-1$ (with multiplicity $n-1$). Let $v_1 = \left(\frac{1}{\sqrt{n}},...,\frac{1}{\sqrt{n}}\right)$, $v_2, ..., v_N$ be an orthonormal basis of $R^N$ composed of eigenvectors of $K_n$, where $v_1$ is the eigenvector associated with $n-1$. Thus, by spectral decomposition of unit vector $e_i = \frac{1}{\sqrt{n}} v_1 + \sum_{j=2}^{n} v_j^i v_j$, we obtain $1 = \|e_i\|^2 = \frac{1}{n} + \sum_{j=2}^{n} \left(v_j^i\right)^2$.

Therefore, we deduce

$$\frac{\mu_k(i)}{k!} = \frac{\langle \mathbf{A}^k e_i, e_i \rangle}{k!} = \frac{1}{n}\left(\frac{(n-1)^k}{k!} + (n-1)\frac{(-1)^k}{k!}\right). \tag{9}$$

Hence, $SC(i) = \frac{1}{n}\left[e^{n-1} + \frac{n-1}{e}\right]$.

**IV. APPLICATIONS TO ARTIFICIAL NETWORKS**

In this section, we present several tests of our centrality measure in "artificial" regular graphs, and we compare it with other centrality measures. We selected regular graphs as a challenging set of graphs because their nodes have identical $DC$ and $EC$. Graph (a) in Fig. 1 also has identical $CC$ for all nodes (normalized $CC$ = 63.636). However, nodes are grouped into three different groups according to $BC$: $\{1,2,8\}$ $BC$ = 9.529, $\{3,5,7\}$ $BC$ = 11.111 and $\{4,6\}$ $BC$ = 7.143. The same clustering is obtained by $SC$ but follows a different order: $\{1,2,8\}$ $SC$ = 3.902, $\{3,5,7\}$ $SC$ = 3.638 and $\{4,6\}$ $SC$ = 3.705. This order is expected in accordance with the number of times each node takes part in the small subgraphs, e.g., triangles and squares, as given in Fig. 1.

Graph (b) in Fig. 1 represents a more challenging example, as it has identical *DC*, *CC*, *EC* and *BC* for all nodes of the graph, and every node participates in the same number of triangles. However, *SC* is able to differentiate nodes $\{1,3,5,6,8\}$ (SC = 45.651) from nodes $\{2,4,7,9\}$ (*SC* = 45.696) following the trend marked by the number of squares in which every node participates; i.e., 44 for nodes in the first group and 45 for nodes in the second. Despite this difference is of only one it clearly indicates that both groups of nodes are different respect to their participation in subgraphs. The difference in the number of other subgraphs (not calculated) could be greater for both graphs, but our objective is to show that different groups of nodes (according to their participation in subgraphs) are differentiated by *SC*, which is clearly observed for the examples given below.

We have calculated *SC* for 210 regular graphs. The number of vertices in the graphs ranged from six to ten, and the degrees of the vertices ranged from three to seven. In all these cases, we have found that for graphs whose nodes all have identical *SC*, all nodes also have identical values of *DC*, *BC*, *CC* and *EC*. However, we have found several examples in which *SC* differentiates nodes even when the other centrality measures are identical. In other words, we have empirically observed that of all centrality measures tested, *SC* had the greatest discriminative power. These characteristics are independent of the size of the graph analyzed and they are straightforwardly generalized for larger regular networks. However, we have not been able to prove this result mathematically for the general case and we propose it in the form of a conjecture:

**Conjecture**: *Let G be a graph having identical subgraph centrality for all nodes. Then the degree, closeness, eigenvector and betweenness centralities are also identical for all nodes.*

## V. APPLICATIONS TO REAL-WORLD NETWORKS

We explored the characteristics of our network subgraph centrality in several kinds of real-world network, including: (i–ii) two protein–protein interaction networks (PINs), one of the yeast

*Saccharomyces cerevisiae* (PIN-1) compiled by Bu et al. [32] on data obtained by von Mering et al. [33] by assessing a total of 80,000 interactions among 5400 proteins assigning each interaction a confidence value. Bu et al. [32] focused on 11,855 interactions between 2617 proteins with high and medium confidence in order to reduce the influence of false positives. The PIN of the bacterium *Helicobacter pylori* (PIN-2) obtained from the Database of Interacting Proteins [34]; (iii–iv) two vocabulary networks in which nodes represent words taken from a dictionary. A directed link from a word to another exists if the first word is used in the definition of the second one. One of these networks is built using the Roget's Thesaurus of English (Roget) [35], and the other is built using the Online Dictionary of Library and Information Science (ODLIS) [36]; (v) a scientific collaboration network in the field of computational geometry compiled from the Computational Geometry Database, version of February 2002 [37] where nodes represent scientists, and two nodes are connected if the corresponding authors wrote a paper together; (vi) a citation network of papers published in the Proceedings of Graph Drawing in the period 1994–2000 [38] where nodes are papers and two nodes are connected if one paper cites another; (vii–viii) the Internet at the autonomous systems (AS) level as of September 1997 and of April 1998 analyzed by Faloutsos et al. [6]. Although some of these relationships are inherently directed, we have ignored direction and consider networks to be undirected for the current analysis. On the other hand, in order to make appropriate comparisons between SC and the other centrality measures, we studied only the main component of these networks owing to the fact that some of the centrality measures cannot be defined for nonconnected graphs. Datasets were collected from the European Project COSIN (http://www.cosin.org/) and from Pajek program datasets (http://vlado.fmf.uni-lj.si/pub/networks/data/).

## VI. COMPARISON TO OTHER CENTRALITY MEASURES

It has been previously shown that strong correlations exist among different centrality measures [39]. This is not surprising because these measures are defined so as to account for the notion of centrality of the nodes in the graph. For instance, nodes with large degrees show in general short average distance to the other nodes in the network, which produces high correlations between node degrees and various measures of centrality. Nodes with large degrees are also expected to participate in large amounts of subgraphs, such as simply connected triplets, triangles, squares and so forth. Consequently, we have observed that, in general, subgraph centrality yields the highest rank orders for those nodes of largest degrees in the network, despite the fact that both measures disagree very significantly for the majority of other nodes (graphics not shown). In the next section, we will analyze the ranking of nodes in more detail.

A global characterization of the network can be carried out by mean of the average subgraph centrality, $\langle SC \rangle$. It has been recommended that the use of centralization instead of centrality is more appropriate for these sort of global measures [8]. An analytical expression for $\langle SC \rangle$ can be obtained using a procedure analogous to that described for proving the previous theorem, showing that $\langle SC \rangle$ depends only on the eigenvalues and size of the adjacency matrix of the network:

$$\langle SC \rangle = \frac{1}{N} \sum_{i=1}^{N} SC(i) = \frac{1}{N} \sum_{i=1}^{N} e^{\lambda_i} \qquad (10)$$

In Table 2 we give the values of $\langle SC \rangle$ as well as the other centralization measures, i.e., average degree $\langle DC \rangle$, average betweenness $\langle BC \rangle$, average closeness $\langle CC \rangle$ and average EC $\langle EC \rangle$, as well as the average clustering coefficient for the whole network, C. We also give the squared correlation coefficients, $R^2$, for the linear regression between the corresponding centralization measure and $\langle SC \rangle$. As we can see in Table 2 $\langle SC \rangle$ is not linearly related to any of the other centralization measures ($R^2 < 0.5$). The only significant relation is obtained between $\langle DC \rangle$ and $\langle SC \rangle$, which indicates that as an average the nodes with larger degrees in the network are also those which participates in a higher number of subgraphs.

Insert Table 2 about here.

**VII. RANKING OF NODES**

One of the most distinctive characteristics of centrality measures is their ability to rank nodes in a network according to the topological features that they account for. It is clear that *DC* takes into account the immediate effect that the closest nodes produce on the corresponding vertex. Our *SC* measure takes into account not only the immediate effects of the closest nodes but also the long-range effects "transmitted" through the participation of a node in all subgraphs existing in the network, giving more weights to shorter subgraphs. Despite these differences, there were several cases in which the ranking of the most central nodes in a network showed great resemblance in both measures. For instance, in the top-10 rankings produced by *DC* and *SC* of the words in the Roget Thesaurus of English, there are seven words that coincide. Eight words in the ODLIS network, seven authors in the Computational Geometry collaboration network and seven nodes in Internet-1997 also coincide for both rankings. In the PIN-1 the number of proteins that coincide in the top-10 rankings is only two, and in PIN-2 there are five. In spite of these coincidences, the exact ranking of the most central nodes differs in order. While "indication" and "deterioration" are the most connected words in Roget, "inutility" and "neglect" are the most central according to SC. L. J. Guibas is the most connected author in the collaboration network of Computational Geometry with 102 coauthors and P. K. Agarwal is the second with 98 coauthors. However, Agarwal is ranked as the most central author according to SC, while Guibas is second. This situation is repeated several times in most of the networks analyzed.

In order to understand the main differences in the orders imposed by these two centrality measures, we have selected an example from the collaboration network of Computational Geometry authors. We selected at random two authors with the same degree and different subgraph centrality (see Fig. 2): Timothy M. Y. Chan and S. L. Abrams, both having $DC = 10$, but having $SC = 8.09 \cdot 10^9$ and $SC = 974.47$, respectively. Despite both authors' having the same number of coauthors, Chan is connected to five of the hubs of this collaboration network: Agarwal (98), Snoeyink (91), Sharir (87), Tamassia (79) and Yap (76) ($DC$ are given in parenthesis). However, Abrams is connected to authors having lower numbers of coworkers; e.g., Patrikalakis has 31 coauthors and the rest have only five to 16 collaborators. This simple difference means that Chan is separated from 623 other authors by a distance of only two; i.e., simply connected triplets, while this number is significantly lower for Abrams, i.e., only 116. The risk that Chan is "infected" with an idea circulating among the authors in this field of research is much higher than the risk with Abrams. This difference is accounted for the subgraph centrality.

Insert Fig. 2 about here.

A similar analysis can be realized for nodes having degree one in a network. According to $DC$, these are the less central nodes of the network. However, we can rank them by $SC$ to see whether one is more or less central. Of all the words in the Roget Thesaurus with degree one, "mart" is ranked by $SC$ as the most central and "sensualist" as the least central. While "mart" is connected to "store", a hub connected to 20 other words, "sensualist" is only connected to "libertine", which is connected only to "impurity", a word linked only to two other words: "purity" and "uncleanness".

**VIII. SUBGRAPH CENTRALITY AND PROTEIN LETHALITY**

In order to investigate the consequences of the differences in the ranking of nodes in real-life scenarios, we have selected the lethality of proteins in *S. cereviciae* (PIN-1). Jeong et al. [40] have shown that the likelihood that removal of a protein from the yeast proteome will prove lethal correlates with the number of interactions that the protein has; i.e., its node degree. We first ranked all proteins in PIN-1 according to both *DC* and *SC*, and then counted the cumulative number of lethal proteins in the first n proteins of the ranking, with an increasing step of 10. For instance, we counted the number of lethal proteins in the first 10 proteins in each ranking, then in the first 20, and so forth. In Fig. 3, we give the general trends for the first 300 proteins in both rankings based on *DC* and *SC*. It can be seen that the ranking introduced by *SC* contains more essential proteins than that introduced by the number of interactions that a protein has. For the first 300 proteins, for example, the number of essential proteins according to *SC* is 148, while according to *DC* it is only 135.

Insert Fig. 3 about here.

In order to understand these differences, we must first investigate which topological features determine the differences in the ranking of proteins according to each centrality measure. The most central proteins according to *DC* are YPR110C and YIL035C, which are transcription proteins, both with 64 interactions. According to *SC*, the most central protein is the transcription protein YNL061W, which has only 48 interactions. However, YNL061W participates in 162 triangles, while the most connected proteins (YPR110C and YIL035C) participate in 52 and 120 triangles, respectively. If we consider the top 10 proteins according to *SC*, the average number of triangles in which each protein participates is 127, while this average is only 57 for the top 10 proteins in the *DC* ranking. Our centrality measure takes into account not only the number of triangles but also the number of simply connected triplets, the number of squares, and other subgraphs in which a node participates. These subgraphs, particularly triangles and squares, can play an important role in understanding the evolution of the protein–protein interaction network [21, 22]. According to the coupled duplication-divergence model of evolution after gene duplication, both of the expressed proteins will have the same interactions [41]. In this model, it is proposed that both duplicate genes are subject to degenerative mutations, losing some functions but jointly retaining the full set of functions present in the ancestral gene. More recently, van Noort et al. [42] have reproduced the scale-free and small-world characteristics of the yeast co-expression network using a similar model, based on the simple neutralist´s model, which consists of co-duplication of genes with their transcription factor binding sites (TFBSs), deletion and duplication of individual TFBSs, and gene loss [42]. Among the effects manifested by these models on the topology of the PIN is the tendency to generate bi-connected triplets and quadruples of nodes; i.e., triangles and squares. Triangles are formed among the duplicating genes and any neighbor of the parent gene, and squares are formed analogously between duplicating genes and any pair of neighbors of the parent gene. These structural features characterizing the topology of the PINs are appropriately measured by the subgraph centrality, which counts the number of weighted subgraphs in which a node of the network participates, giving higher weights to smaller subgraphs. We therefore conclude that our finding concerning the centrality–lethality relation in the yeast PIN is a consequence of the fact that indispensability of a given protein in the PIN is more a consequence of its imbrications in certain structural motifs, such as triangles and squares, than of its connectivity.

**IX. SCALING PROPERTIES**

In a general classification of small-world networks, Amaral et al. [43] have presented empirical evidence for the occurrence of three structural classes. According to the cumulative distribution of vertex degrees, they found: (i) scale-free networks, characterized by a connectivity distribution with a tail that decays as a power law; (ii) truncated scale-free networks, characterized by a connectivity distribution that has a power-law regime followed by a sharp cutoff of the tail; and

(iii) single-scale networks, characterized by a connectivity distribution with a fast decaying tail. Power-law distributions have also been observed for the betweenness centrality in several types of network, which have been used to classify scale-free networks [44].

In the following, we use cumulative rather than density distribution of both *DC* and *SC*, based on the work of Amaral et al. [43] and other evidence for its advantages in small, noisy data sets [39]. All eight networks studied displayed a cumulative subgraph centrality distribution that corresponded with scale-free characteristics. In Fig. 4, we illustrate the linear-log plots of the cumulative distributions of *SC* (left) and *DC* (right) for the eight networks. Interestingly, the PIN of *S. cereviciae* does not display scale-free degree distribution but rather corresponds with a broad-scale network, in which a power-law regime is followed by a large tail that decays according to an exponential or Gaussian law. We have investigated this distribution in detail for this network and observed a power-law distribution for the region of lower degree, with a squared correlation coefficient greater than 0.98. Similar behavior was found by Amaral et al. for the movie actor network, first reported as scale free [19] and then later found to display truncated scale-free characteristics. Recently, Newman has reported that three bibliographic networks in the fields of biology, physics, and mathematics do not follow power laws, but probably display broad-scale behavior [45].

Insert Fig. 4 about here.

The vocabulary network of the Roget Thesaurus and the citation network of Graph Drawing Proceedings are both single-scale networks following a connectivity distribution with an exponential or Gaussian decaying tail. However, they both show clear scale-free subgraph centrality distributions. The scale-free characteristics of the *SC* distribution can be explained as follows. *SC* measures the number of times a node participates in all subgraphs in the network, giving more weight to smaller subgraphs. Consequently, nodes with high *SC* participate in a high number of small subgraphs, such as connected triplets, triangles, squares, etc. The frequency of these nodes in the network is significantly lower than that of nodes participating in a small number of subgraphs or participating only in large subgraphs from which a fat tail distribution results. These scale-free behaviors of the *SC* distribution are not expected to be universally followed for all kinds of network. In fact, we have found exponential decay distributions for *SC* in some networks, such as food webs.

## X. CONCLUSIONS

We have proposed a new centrality measure for the nodes of a network, based on spectral properties, which shows interesting and desirable properties. It characterizes nodes according to their participation in structural subgraphs in the network, giving higher weights to the smaller subgraphs that can be involved in network motifs. This centrality has been tested in artificial networks, showing that it is more discriminative than degree, betweenness, closeness or eigenvector centrality for the nodes of a network. In real-world complex networks, the subgraph centrality does not show strong correlation with other centrality measures, and it gives a distinctly different ranking of nodes. In the networks studied here, subgraph centrality displays a power-law distribution even in cases in which degree centrality does not display a scale-free distribution.

**ACKNOWLEDGEMENTS**

EE thanks "Ramon y Cajal" Program for partial financial support. The critics and suggestions of two anonymous reviewers contribute to improve the quality of this presentation.

**Figure captions**

FIG. 1. Examples of regular graphs with nodes distinguished by subgraph centrality but not by other centrality measures. All nodes in graph (a) have identical *DC*, *CC* and *EC* but are distinguished by *BC* and *SC*. The numbers of triangles and squares are given as an ordered pair in parentheses. In graph (b), all nodes have identical *DC*, *CC*, BC, and *EC* but are differentiated by *SC*.

FIG. 2. Subgraphs of the collaboration network in Computational Geometry for two author with the same degree centrality but different subgraph centrality: Timothy M. Y. Chan and S. L. Abrams and all their coworkers.

FIG. 3. The number of essential proteins in the PIN of *S. cereviciae* according to the ranking of nodes produced by *DC* (red) and *SC* (blue).

FIG. 4. Linear-log plot of the cumulative distribution of *SC* (left) and *DC* (right) in eight complex networks.

[1] S. H. Strogatz, Nature (London) **410**, 268 (2000).

[2] R. Albert and A.-L. Barabási, Rev. Mod. Phys. **74**, 47 (2002).

[3] S. N. Dorogovtsev and J. F. F. Mendes, Adv. Phys. **51**, 1079 (2002).

[4] M. E. J. Newman, SIAM Rev. **45**, 167 (2003).

[5] A.-L. Barabási and Z. N. Oltvai, Nature Rev. Genet. **5**, 101 (2004).

[6] M. Faloutsos, P. Faloutsos and C. Faloutsos, Comp. Comm. Rev. **29**, 251 (1999).

[7] R. Albert, H. Jeong and A.-L. Barabási, Nature (London) **401**, 130 (1999).

[8] S. Wasserman and K. Faust, *Social Network Analysis* (Cambridge Univ. Press, Cambridge, UK, 1994).

[9] F. Liljeros, C. R. Edling, L. A. N. Amaral, H. E. Stanley and Y. Åberg, Nature (London) **411**, 907 (2001).

[10] A. Schneeberger, C. H. Mercer, S. A. J. Gregson, N. M. Ferguson, C. A. Nyamukapa, R. M. Anderson, A. Johnson G. P. Garnett, Sex. Transm. Dis. **31**, 380 (2004).

[11] M. E. J. Newman, Proc. Natl. Acad. Sci. USA **98**, 404 (2001).

[12] R. Ferrer i Cancho and R. V. Solé, Proc. R. Soc. Lond. B **268**, 2261 (2001).

[13] M. Sigman and G. A. Cecchi, Proc. Natl. Acad. Sci. USA **99**, 1742 (2002).

[14] D. J. Watts and S. H. Strogatz, Nature (London) **393**, 440 (1998).

[15] R. J. Williams and N. Martinez, Nature (London) **404**, 180 (2000).

[16] H. Jeong, B. Tombor, R. Albert, Z. N. Oltvai and A. L. Barabási, Nature (London) **407**, 651 (2000).

[17] S. Wuchty, Mol. Biol. Evol. **18**, 1694 (2001).

[18] D. J. Watts, *Small World: The Dynamics of Networks Between Order and Randomness* (Princeton Univ. Press, Princeton, NJ, 1999).

[19] A.-L. Barabási and R. Albert, Science **286**, 509 (1999).

[20] M. Girvan and M. E. J. Newman, Proc. Natl. Acad. Sci. USA **99**, 7821 (2002).

[21] R. Milo, S. Shen-Orr, S. Itzkovitz, N. Kashtan, D. Chklovskii and U. Alon, Science **298**, 824 (2002).

[22] E. Yeger-Lotem, S. Sattath, N. Kashtan, S. Itzkovitz, R. Milo, R. Y. Pinter, U. Alon and H. Margalit, Proc. Natl. Acad. Sci. USA **101**, 5934 (2004).

[23] L. C. Freeman, Social Networks **1**, 215 (1979).

[24] L. C. Freeman, Sociometry **40**, 35 (1977).

[25] P. Bonacich, J. Math. Sociol. **2**, 113 (1972).

[26] P. Bonacich, Am. J. Sociol. **92**, 1170 (1987).

[27] D. Cvetkovic, P. Rowlinson and S. Simic, *Eigenspaces of Graphs* (Cambridge Univ. Press, Cambridge, UK, 1997).

[28] M. E. J. Newman, Phys. Rev. E **70**, 056131 (2004).

[29] I. Farkas, I. Derényi, H. Jeong, Z. Néda, Z. N. Oltvai, E. Ravasz, A. Schubert, A.-L. Barabási and T. Vicsek, Physica A **314**, 25 (2002).

[30] I. Farkas, I. Derényi, A.-L. Barabási and T. Vicsek, Phys. Rev. E. **64**, 026704 (2001).

[31] B. Ruhnau, Social Networks **22**, 357 (2000).

[32] D. Bu, Y. Zhao, L. Cai, H. Xue, X. Zhu, H. Lu, J. Zhang, S. Sun, L. Ling, N. Zhang, G. Li and R. Chen, Nucl. Acids Res. **31**, 2443 (2003).

[33] C. von Mering, R. Krause, B. Snel, M. Cornell, S. G. Oliver, S. Fields P. Bork, Nature (London) **417**, 399 (2002).

[34] J.-C. Rain, L. Selig, H. De Reuse, V. Battaglia, C. Reverdy, S. Simon, G. Lenzen, F. Petel, J. Wojcik, V. Schächter, Y. Chemama, A. Labigne and P. Legrain, Nature (London) **409**, 211 (2001).

[35] Thesaurus (2002) Roget´s Thesaurus of English Words and Phrases. Project Gutenberg. http://gutenberg.net/etext/22

[36] ODLIS (2002) Online Dictionary of Library and Information Science. http://vax.wcsu.edu/library/odlis.html

[37] B. Jones, *Computational Geometry Database*, February, 2002. http://compgeom.cs.edu/~jeffe/compgeom/biblios.html

[38] V. Batagelj and A. Mrvar, *Graph Drawing Contest 2001*. http://vlado.fmf.uni-lj.si/pub/GD/GD01.htm

[39] S. Wuchty and P. F. Stadler, J. Theor. Biol. **223**, 45 (2003).

[40] H. Jeong, S. P. Mason, A.-L. Barabási and Z. N. Oltvai, Nature (London) **411**, 41 (2001).

[41] A. Force, M. Lynch, F. B. Pickett, A. Amores, Y. Yan and J. Postlethmait, Genetics **151**, 1531 (1999).

[42] V. van Noort, B. Snel and M. A. Huynen, EMBO Reports **5**, 1 (2004).

[43] L. A. N. Amaral, A. Scala, M. Barthélemy and H. E. Stanley, Proc. Natl. Acad. Sci. USA **97**, 11149 (2000).

[44] K.-I. Goh, E. Oh, H. Jeong, B. Kahng and D. Kim, Proc. Natl. Acad. Sci. USA **99**, 12583 (2002).

[45] M. E. J. Newman, Proc. Natl. Acad. Sci. USA **101**, 5200 (2004).

(a)

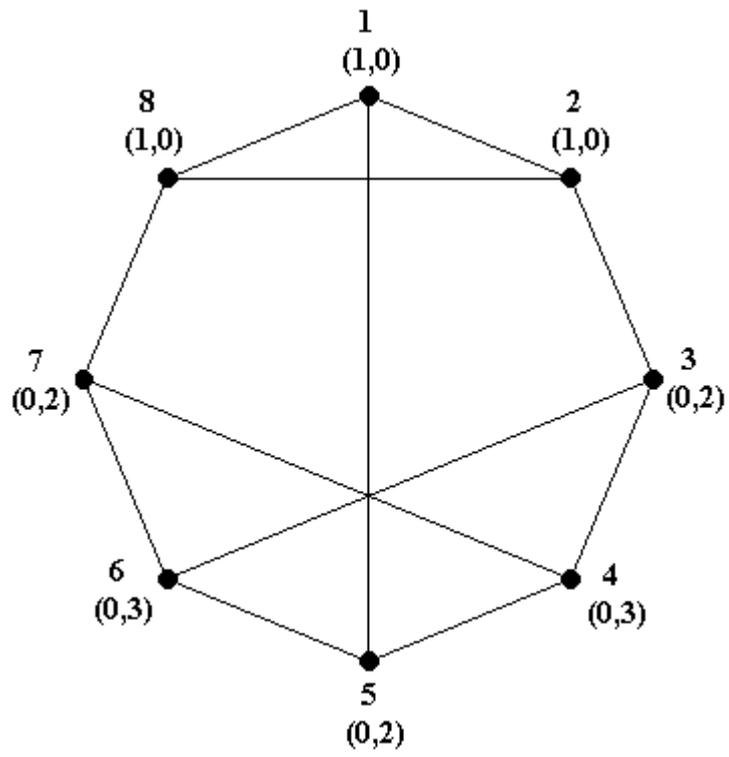

(b)

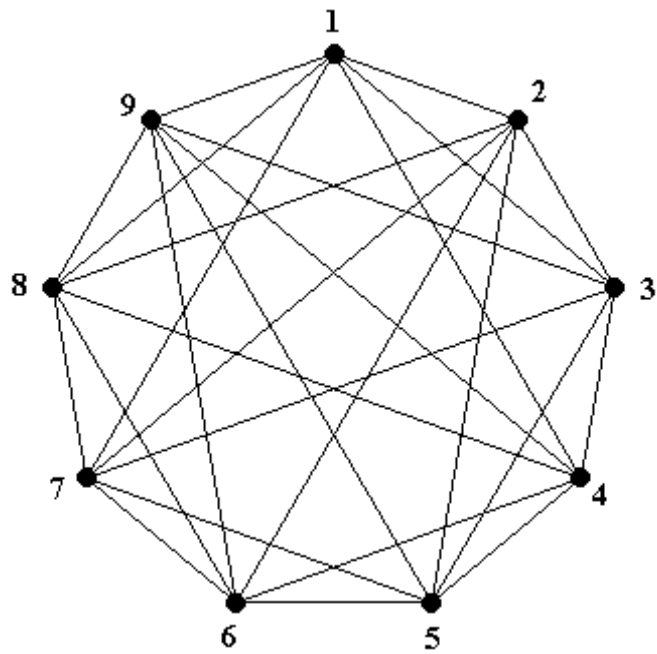

**Fig. 2**

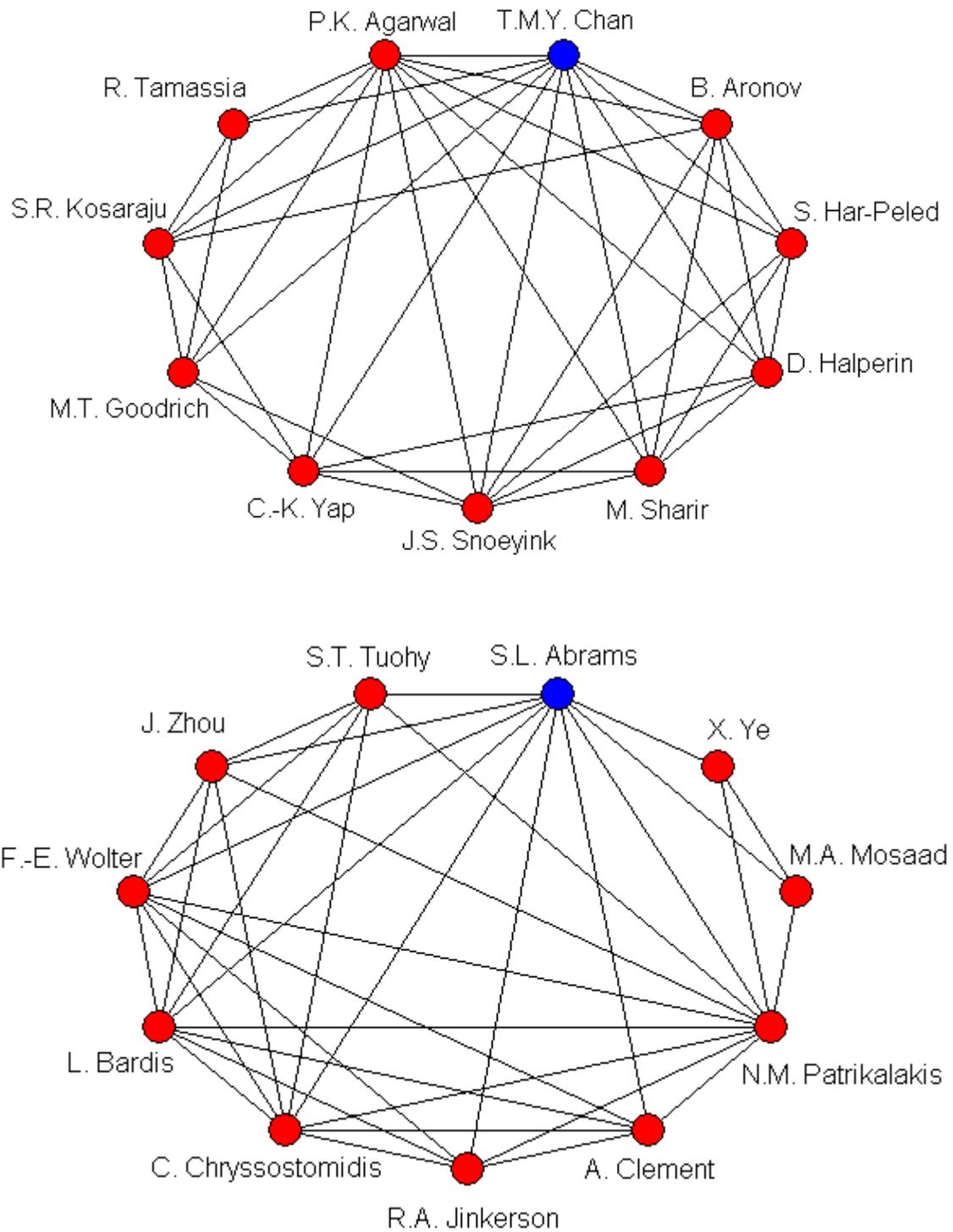

**Fig. 3**

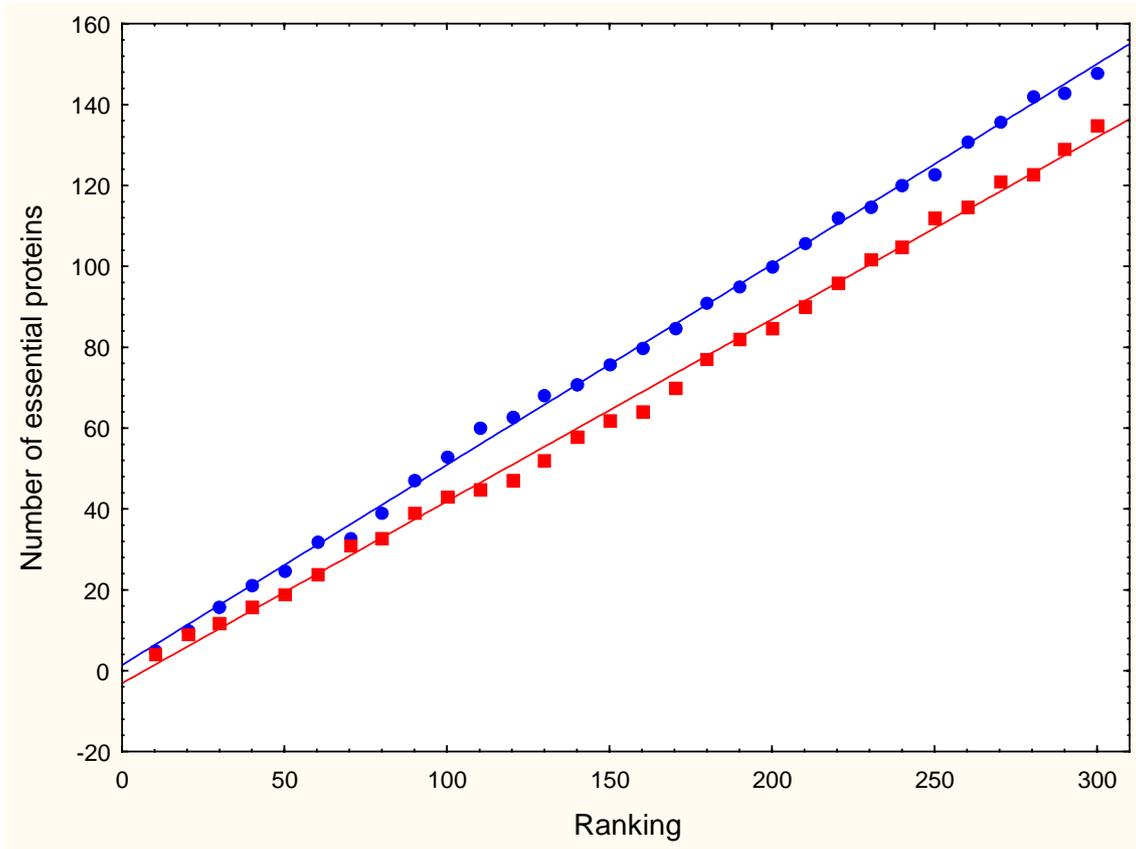

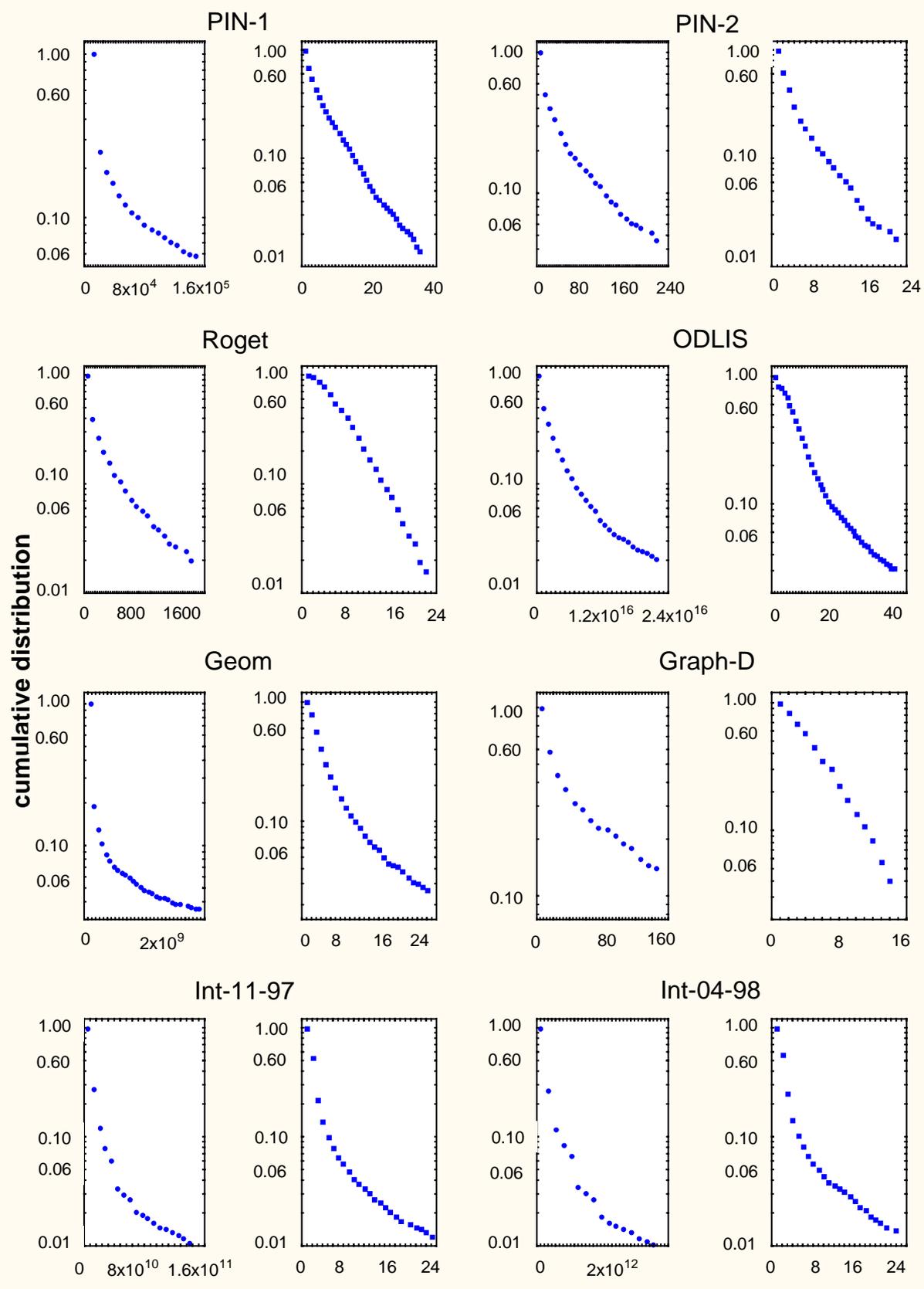

Table 1. Illustration of the relationship between closed walks (trivial and non-trivial) of length four and the subgraphs associated to them.

| Type | Closed Walk | Subgraph |
|---|---|---|
| Trivial | 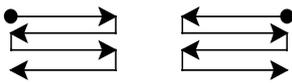 | 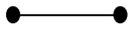 |
| Trivial | 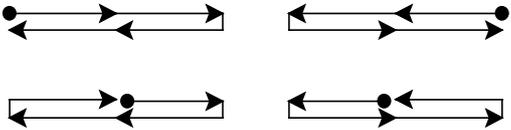 | 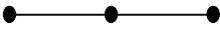 |
| Non trivial | 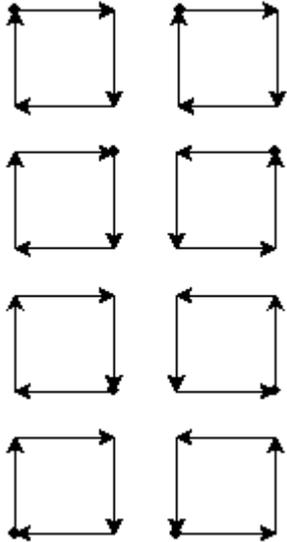 | 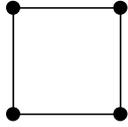 |

Table 2. Summary of results of eight real-world complex networks.

| Network | Nodes | Links | $\langle DC \rangle$ | $\langle BC \rangle$ | $\langle CC \rangle$ | $\langle EC \rangle$ | $C$ | $\langle SC \rangle$ |
|---|---|---|---|---|---|---|---|---|
| PIN-1 | 2224 | 6608 | 5.94 | 3752.7 | 23.3 | 0.0078 | 0.200 | 87269.5 |
| PIN-2 | 710 | 1396 | 3.93 | 1117.5 | 24.5 | 0.0219 | 0.025 | 64.7 |
| Roget | 994 | 3640 | 7.32 | 1526.9 | 24.9 | 0.0209 | 0.162 | 239.4 |
| ODLIS | 2898 | 16376 | 11.30 | 3142.9 | 32.1 | 0.0107 | 0.351 | $5.3 \times 10^{15}$ |
| Geom | 3621 | 9461 | 5.22 | 7811.2 | 19.5 | 0.0047 | 0.679 | $1.1 \times 10^{9}$ |
| GD | 249 | 635 | 5.10 | 390.6 | 24.8 | 0.0378 | 0.287 | 64.3 |
| Int-97 | 3015 | 5156 | 3.42 | 4161.6 | 27.3 | 0.0082 | 0.348 | $2.05 \times 10^{10}$ |
| Int-98 | 3522 | 6324 | 3.59 | 4870.8 | 27.3 | 0.0076 | 0.340 | $4.04 \times 10^{11}$ |
| $R^2$ | | | 0.748 | 0.001 | 0.543 | 0.023 | 0.012 | |

*DC*, degree centrality; *CC*, normalized closeness centrality; *BC*, betweenness centrality; *EC*, eigenvector centrality; *SC*, subgraph centrality; *C*, Watts-Strogatz clustering coefficient [14], $\langle \cdots \rangle$ symbol is used for average values for all nodes of the network. $R^2$ is the square correlation coefficient of the linear regression between the corresponding centrality measure and $\langle SC \rangle$.